%% file: paper.tex
\documentclass[runningheads]{llncs}

\input{pack}

\begin{document}

\title{Identifying Asymptomatic Nodes in Network Epidemics using Graph Neural Networks\thanks{Paper presented in the 35th Brazilian Conference on Intelligent Systems (BRACIS)}}
\titlerunning{Identifying Asymptomatic Nodes in Network Epidemics using GNNs}

\author{Conrado Catarcione Pinto \and
Amanda Camacho Novaes de Oliveira \and \\
Rodrigo Sapienza Luna \and 
Daniel Ratton Figueiredo}

\authorrunning{C. Catarcione Pinto et al.}

\institute{Computer Sciences and Systems Engineering Department, \\ Universidade Federal do Rio de Janeiro (UFRJ)\\
Rio de Janeiro, RJ, Brazil \\ \email{\{conrado, amandacno, rluna, daniel\}@cos.ufrj.br}}

\maketitle              
\begin{abstract}
Infected individuals in some epidemics can remain asymptomatic while still carrying and transmitting the infection. These individuals contribute to the spread of the epidemic and pose a significant challenge to public health policies. Identifying asymptomatic individuals is critical for measuring and controlling an epidemic, but periodic and widespread testing of healthy individuals is often too costly. This work tackles the problem of identifying asymptomatic individuals considering a classic SI (Susceptible-Infected) network epidemic model where a fraction of the infected nodes are not observed as infected (i.e., their observed state is identical to susceptible nodes). In order to classify healthy nodes as asymptomatic or susceptible, a Graph Neural Network (GNN) model with supervised learning is adopted where a set of node features are built from the network with observed infected nodes. The approach is evaluated across different network models, network sizes, and fraction of observed infections. Results indicate that the proposed methodology is robust across different scenarios, accurately identifying asymptomatic nodes while also generalizing to different network sizes and fraction of observed infections.

\end{abstract}



\input{introduction}

\input{literatura}

\input{modelo}

\input{metodo}

\input{resultados}

\input{conclusao}

\begin{credits}
\subsubsection{\ackname} This work received financial support through research grants from CNPq (310742/2023-4), FAPERJ (E-26/200.483/2023 and E-26/202.517/2024) and CAPES (PROEX).

\subsubsection{\discintname}
 The authors have no competing interests to declare that are relevant to the content of this article. 
\end{credits}

\bibliographystyle{splncs04}
\bibliography{paper}

\end{document}

%% file: pack.tex
\usepackage[T1]{fontenc}
\usepackage{graphicx}
\usepackage{hyperref}
\usepackage{amsmath}
\usepackage{amsfonts}
\usepackage{dsfont}
\usepackage{csquotes}
\usepackage{subfig}
\usepackage{soul}
\usepackage{multirow}	
\usepackage{booktabs}  
\usepackage[dvipsnames]{xcolor}
\usepackage{nicefrac}

\usepackage[colorinlistoftodos, color=yellow]{todonotes}

\newcommand{\mil}{\,\text{k}}

\newcommand{\meanRel}[2]{\text{mean}_{#2} \left( {#1} \right)}
\newcommand{\stdRel}[2]{\text{std}_{#2} \left( {#1} \right)}

\colorlet{tableLines}{BlueViolet} 
\colorlet{tableLines2}{Brown!60!Black} 

\graphicspath{ {imagens} }

\hypersetup{
    allcolors=MidnightBlue, 
    colorlinks=true,
}
\allowdisplaybreaks

%% file: introduction.tex
\section{Introduction}
\label{sec:intro}
	

The fast spread of a virus or other biological agent through a population, known as an epidemic, can often lead to serious health and economic crises such as the COVID-19 pandemic~\cite{Kaye_2021,Lu_2021}. Identifying infected individuals is a cornerstone towards measuring the spread of an epidemic and planning public policies that can mitigate it. For example, these individuals can be quarantined in order to avoid contact with others. However, this requires correctly determining which people have, in fact, been contaminated.

Individuals who have symptoms of the related disease can often be identified and potentially tested for more accurate identification. However, some infected individuals do not exhibit any symptoms (known as {\em asymptomatic}) but still have the potential to contaminate others and contribute to the epidemic's spread \cite{Inui}. While asymptomatic individuals can often be identified as infected through testing, determining which healthy individuals to test is a challenge. In particular, periodically checking healthy individuals in a large population is often too costly to be practical~\cite{Segui}. Thus, mechanisms that can accurately identify asymptomatic individuals in a large population prior to diagnosis become important. This work tackles this problem in the context of network epidemic models. 

The classic SI (Susceptible-Infected) network epidemic model considers an epidemic that spreads through a network where susceptible nodes (S) can become infected by neighbouring infected nodes (I). In this model, nodes are either susceptible (S) or infected (I), and once infected, the node remains infected indefinitely (see Section \ref{sec:modelo}). This model represents some common viruses behaviours, such as HIV and certain types of hepatitis. 

Consider an SI network epidemic where a fraction of the infected nodes are not observed as infected, meaning they are asymptomatic and appear identical to susceptible nodes. Can such nodes be identified, given the network and the observed infected nodes? This question has been recently tackled using a network metric, the observed betweenness, 
that was designed so that nodes with higher observed betweenness are more likely to be asymptomatic~\cite{Catarcione_2024}. Although relatively simple, the approach was shown to be effective against other baselines. 

This work tackles the problem using a classic Graph Neural Network (GNN) model that classifies apparent healthy nodes as infected (asymptomatic) or susceptible using supervised learning. 
Given that the input is a network and the epidemic state (susceptible/asymptomatic or infected) of the nodes, a natural model choice for node classification is the GNN.
However, this simple (binary) state can be augmented with richer features extracted from the input (network and set of observed infected nodes). In particular, this work adopts a set of network epidemic features, including the observed betweenness, recently proposed for identifying asymptomatic nodes~\cite{Catarcione_2024}. These features are used as input for the GNN model, so as to enable it to identify the asymptomatic nodes. 
Features for tackling the problem of identifying the sources of an epidemic
~\cite{Haddad_2023} were also applied here, adding extra inputs to help the GNN find asymptomatic nodes.

This proposed framework is evaluated in various scenarios, including two network models (BA and WS), different network sizes, and different fractions of asymptomatic nodes~(see Section \ref{sec:result}). Note that the network epidemic and infected observation models can be jointly used to generate datasets for training (supervised learning) and testing the GNN model, enabling a more comprehensive evaluation. In comparison with prior work, the results obtained are very promising and can surpass the observed betweenness baseline. In particular, the model can generalise well to larger networks and different fractions of asymptomatic nodes. However, the effectiveness of the GNN in BA networks is not always superior to the observed betweenness baseline, indicating room for possible improvements in the framework. 

The remainder of this paper is organised as follows. Section~\ref{sec:lit} presents a brief discussion of the related literature. 
Section~\ref{sec:modelo} presents the network epidemic model used throughout this work and the subsequent infected observation model. 
Section~\ref{sec:metodo} describes the features adopted by the proposed GNN. 
The evaluation metrics, scenarios, and results are presented and discussed in Section~\ref{sec:result}. 
Section~\ref{sec:conc} concludes the paper with brief remarks.

%% file: literatura.tex
\section{Related Work}
\label{sec:lit}
    Graph Neural Networks (GNNs) have attracted significant attention in recent epidemic research due to their ability to model complex relational data. As outlined by Liu et al. \cite{surveyGNNepidemia}, these machine learning models support a wide range of applications. One key area of focus is epidemic source detection, where Graph Convolutional Networks (GCN) have been employed to identify both single-source \cite{Sha_2021,Shah_2020} and multi-source outbreaks \cite{Haddad_2023}. GNNs have also been applied to infection risk surveillance in geographical locations \cite{Han_2023,Yu_2023} and within hospitals to monitor patient risk \cite{Gouareb_2023}. Additionally, GNNs have also been explored as tools for predicting the impact of infected individuals on future infections \cite{Song_2023}.




    
    The problem of identifying asymptomatic individuals within a network epidemic has been recently addressed by Chen et al. \cite{Chen_2023}.
    Their work considers an epidemic model where susceptible nodes first become asymptomatic upon infection, but eventually transition to a symptomatic state, and finally recover. They propose the TrustRank algorithm to rank nodes by their likelihood of being asymptomatic and evaluate if the single top-ranked node is indeed asymptomatic. 

    Huang et al. \cite{Huang_2023} present a different epidemic model in which susceptible nodes may transition to either an asymptomatic or a presymptomatic state, with the latter eventually progressing to a symptomatic infection. All infection pathways ultimately lead to recovery. They infer asymptomatic nodes by modelling the epidemic state transitions using a Markovian process that incorporates not only network topology but also the duration in different epidemic states. Given the full infection history of symptomatic individuals, the model estimates infection probabilities for the remaining nodes. 
    The performance of their method is compared, and shown to be superior, to a GCN model with trivial node features. In contrast, this present work considers a set of rich node features computed from the input for the GNN model, and also no timing information concerning epidemic states is available (our input is a single snapshot of the epidemic).
    

    Zhang, Tai \& Pei \cite{Zhang_2023} addressed the problem of inferring unobserved infections in a network-based SIR epidemic model using daily testing data to partially observe infected nodes. They propose an ensemble Bayesian inference framework to detect unobserved infections. By combining backward temporal propagation of future observations with cross-ensemble covariability adjustments, their method allows the iterative refinement of infection probability estimates for unobserved nodes. 
    Because it relies on equations that assume a locally tree-like structure, the inference accuracy is reduced in real-world networks with high clustering.
    
    Pinto \& Figueiredo \cite{Catarcione_2024} propose a straightforward method based on network centrality and infection observations on an SI epidemic model. Betweenness centrality is altered to consider exclusively the shortest paths between pairs of nodes that are both observed as infected (observed betweenness).
    Their work is the closest to this paper as both consider the same SI epidemic model and the same network models. However, this work uses a GNN model to classify asymptomatic nodes, utilizing the observed betweenness as one of several input features.
    
    Considering these previous studies, this work aims to improve the application of GNNs in the challenging problem of identifying asymptomatic individuals when no timing information is available.

%% file: modelo.tex
\section{Network Epidemic and Observation Models}
\label{sec:modelo}
To address the asymptomatic detection problem, we adopt a network epidemic model that has become widely established and extensively employed over the past decades \cite{Keeling_2005}. In contrast to classical epidemic models based on systems of differential equations, which capture the dynamics at a population level and fail to account for individual heterogeneity, network epidemic models enable the representation of individual-level contacts and are, therefore, better suited to emulate real-world epidemic processes. 

Alongside this underlying network structure, a compartmental epidemic model is also adopted, in which individuals transition between epidemiological states according to predefined rules. The classic Susceptible-Infected (SI) model is considered here. In this model, individuals are initially healthy - or susceptible to getting infected - and may contract the disease through an infected individual. Once infected, they remain in that state permanently. This study assumes a discrete-time model, meaning that transitions in epidemiological states occur only between successive time steps. 

Let $V$ and $E$ denote the set of nodes and (undirected) edges of the network, respectively, and $N(v)$ the set of neighbours of node $v \in V$.
$S(t)$ and $I(t)$ represent the set of susceptible and infected nodes at time $t \geq 0$, respectively. Since individuals can only belong to one epidemiological state at a time, these sets partition the node set $V$:
\begin{align*}
    S(t) \cup I(t) &= V \\
    S(t) \cap I(t) &= \emptyset,
\end{align*}
for all $t \in \mathbb{N}$.

The epidemic is initialised at time zero by randomly and uniformly selecting a single node to be infected, known as the epidemic source. Afterwards, the epidemic spreads across the network according to a probabilistic transmission rule. In particular, with probability $\beta$, an infected neighbour of a susceptible $v$ node infects node $v$ in the next time step. Thus, the probability that a susceptible node becomes infected at the next time step (by any of its infected neighbours) is given by the complement of the chance it avoids infection from all of its infected neighbours. Specifically, the infection probability is given by:
\begin{equation}
    \mathbb{P} \left[ v \in I(t+1) | v \in S(t) \right] = 1 - \left( 1 - \beta \right)^{r(v, t)},
\end{equation} 
where $\beta \in \left[0,1\right]$ is the infection probability through an edge, and $r(v, t)$ is the number of infected neighbours of node $v$ at time $t$, computed as
\begin{equation}
    r(v,t) = \sum_{u \in N(v)} \mathds{1}(u \in I(t)),
\end{equation}
with $N(v)$ being the set of neighbours of node $u$ and $\mathds{1}$ is the indicator function.

Note that $\beta$ is the single parameter for this probabilistic epidemic model. Moreover, note that in this model, all nodes will eventually become infected (given a large enough time horizon). 

\subsection{Observation model}
After a time $t_h$, the epidemic stops evolving and a snapshot of the network is taken, capturing the infection status of each node. At this point, we introduce a simple probabilistic observation model to distinguish between symptomatic (i.e., observed as infected) and asymptomatic infections (i.e., not observed as infected). Specifically, each infected node in $I(t_h)$ is independently observed with probability $\theta$, which may be interpreted as the probability of symptom manifestation. Thus, the probability that a node is asymptomatic is $1-\theta$. 

We denote by $O(t_h) \subseteq I(t_h)$ the set of observed infected nodes at time $t_h$, corresponding to the snapshot moment. The remaining infected individuals - those not observed - are considered asymptomatic. We define this subset as $A(t_h)$, given by:
\begin{equation}
    A(t_h) = I(t_h) \setminus O(t_h).
\end{equation}

From an observational point of view, these asymptomatic nodes are indistinguishable from the susceptible nodes and are precisely the ones to be identified by the classification algorithm. Thus, the goal is to identify nodes in $A(t_h)$ from the set $A(t_h) \cup S(t_h)$.

%% file: metodo.tex
\section{Features for the GNN Model}
\label{sec:metodo}
        Consider a single snapshot of an SI epidemic at time $t_h$, where all the available information is the network structure (its nodes and edges) and the set of observed infected nodes, $O(t_h)$. This is the input to the problem of identifying the asymptomatic nodes.

        From this information only, it is possible to extract more insights about the epidemic spread, which can be helpful to the AI discriminator. This extra knowledge can be encoded as node features, which are then used as input to the GNN model. In particular, the following eight features were computed and incorporated:
        \begin{itemize}
            \item \textbf{Infection observation.} If the node was observed as infected or not.
            \begin{equation*}
                C_{\textit{o}}(v, t) = \begin{cases}
                    1,& \textit{if } v \in O(t) \\
                    0,& \textit{otherwise}
                \end{cases}
            \end{equation*}
            \item \textbf{Degree.} Number of neighbours of the node.
            \item \textbf{Contact-$k$.} Fraction of infected nodes observed at distance of exactly $k$ from the node \cite{Haddad_2023}, for $k \in \{1,2,3\}$.
            \begin{equation*}
                C_{\textit{c}_{k}}(v, t_h) = \frac{|N_{k}(v) \cap O(t_h)|}{|N_{k}(v)|},
            \end{equation*}
            where $N_{k}(v)$ is the set of nodes at distance exactly $k$ from node $v$.
            \item \textbf{Neighbourhood Contact-2.} Fraction of infected nodes observed with distance 2 or less from the node \cite{Haddad_2023}.
            \begin{equation*}
                C_{\textit{c}_{\leq2}}(v, t_h) = \frac{|N_{\leq 2}(v) \cap O(t_h)|}{|N_{\leq 2}(v)|},
            \end{equation*}
            where $N_{\leq 2}(v)$ is the set of nodes at distance at most 2 from node $v$, i.e., $N_{\leq 2}(v) = N(v) \cup N_{2}(v)$. 
            \item \textbf{Betweenness.} Node's betweenness metric \cite{Freeman_1977}.
            \item \textbf{Observed betweenness.} A variation of the betweenness metric where only the shortest paths between pairs of observed infected nodes are considered. Recall that this metric has been proposed to identify asymptomatic nodes~\cite{Catarcione_2024}.
            \begin{equation*}
               C_{\textit{ob}}(v, t) = \sum_{x, y \in O(t_h); x, y \neq v} \frac{\sigma(x, y | v)}{\sigma(x,y)}, 
            \end{equation*}
            where $\sigma(x,y)$ is the number of shortest paths between nodes $x$ and $y$, and $\sigma(x,y|v)$ is the number of those paths that includes node $v$.
        \end{itemize}        
        
        In order to handle various graph sizes and other variations, all features are normalised before being used as input for the GNN model. The only exception is the first feature, the infection observation, which is a binary value. All other features were normalised to have zero mean and unit variance within a problem instance. That is, for each feature $C_f$ different from the infection observation and each node $v$ of a given problem instance and graph $G$, an input $\overline{C}_f(v)$ is generated as follows:
        \begin{equation}
        	\overline{C}_f (v) = \dfrac{ C_f (v) - \meanRel{C_f}{G} }{ \stdRel{C_f}{G} } \,\,,
        \end{equation}
        where $V_G$ is the set of nodes in $G$, $\meanRel{C_f}{G} = \frac{1}{|V_G|} \sum_{u \in V_G} C_f(u) $ is the average value of feature $f$ within the nodes of graph $G$, and its standard deviation corresponds to $ \stdRel{C_f}{G} = \sqrt{ \frac{1}{|V_G|} \sum_{u \in V_G} \left[ G_f(u) - \meanRel{C_f}{G} \right]^2 }$. 


        After normalisation and model inference, the output is a score between 0 and 1 for each node in the graph, such that larger values indicate that the node is more likely to be asymptomatic, according to the GNN model.


%% file: resultados.tex
\section{Evaluation}
\label{sec:result}

    
    
	
	\subsection{Experimental setup}
	\label{sec:result:setup}

        \subsubsection{Data setup.}
        A dataset of snapshots 
        from epidemics on random networks was generated for the evaluation of GNN's discrimination capacity (in terms of accurately identifying asymptomatic nodes).

        These networks were sampled from two random network models, which have structural properties found in many real-world networks: the Barabási-Albert (BA) \cite{BA} and the Watts-Strogatz (WS) \cite{WS} models. BA networks' degree distribution follows a power law, which is often seen in real networks \cite{BA}, while WS generates networks with the \enquote{small-world} property, which means a few hops are sufficient to reach most network nodes while the clustering coefficient of the network is relatively high \cite{WS}.


        The instance generation process is as follows: once the random network is generated, an SI epidemic is simulated on this network according to the epidemic model described in Section \ref{sec:modelo}. 
        A single node is randomly chosen as the epidemic source, and the infection probability $\beta$ is uniformly chosen from the set $\{0.1, 0.3, 0.5\}$. 
        The epidemic stops when the number of infections reaches 20\% of all the nodes in the network. Thus, $t_h = \arg\min_t \nicefrac{I(t)}{|V_G|} \geq 0.2$.
        Using the observation model, each infected node is observed with probability $\theta$, and five different observation probabilities were considered: $\theta \in \{0.1, 0.25, 0.5, 0.75, 0.9\}$.

        In total, 10 different training datasets were generated, each one with a particular network model (BA or WS) and an observation probability $\theta$. 
        The generated BA networks used parameter $m=4$, and the WS networks used parameters $p=0.3$ and $k=8$. 
        All of them were created with $|V_G| = 3\mil$ nodes, and each dataset contained $1\mil$ instances.
        For each dataset, a separate GNN was trained, totalling 10 different models to be evaluated, each model trained on all $1\mil$ epidemic snapshots of $3\mil$ nodes each (600 of which are infected).

        For the test datasets, all the parameters were kept the same, except for the number of nodes: aside from $3\mil$ nodes, datasets were created using $1\mil$, $6\mil$, and $12\mil$ nodes. In total, $40$ test datasets were generated, summing up to $40\mil$ instances. The instances were all generated with different random seeds, guaranteeing that there is no data leakage between training and testing, and that all training and testing samples are statistically independent.

        The datasets used for training and testing the different scenarios are publicly available at \href{https://doi.org/10.5281/zenodo.15376204}{doi.org/10.5281/zenodo.15376204}.
        

        \subsubsection{Model setup.}
        As for the GNN model, a two-layer GCN \cite{Book_Hamilton_2020,Kipf_2016} was employed, with a ReLU activation in the first layer and a sigmoid activation in the output layer. The model learns the likelihood of each node being infected, based on binary ground-truth labels. Training was performed using the Binary Cross-Entropy loss (thus, supervised learning was performed), and performance analysis (both for training and testing) was restricted to nodes \( v \notin O(t_h) \), i.e., nodes whose infection status was not observed at time \( t_h \). The model was trained for a fixed number of 1000 epochs using the Adam optimizer with a learning rate of \( 10^{-3} \), a batch size of 128, and a hidden layer embedding size of 128. Part of the training dataset was reserved for model selection, used to assess the trained model's performance every 50 epochs. After $1000$ epochs, the best model in the validation dataset was selected as the output model. 
    
        Two evaluation metrics are used: the top-$k$ precision, which measures the fraction of asymptomatic nodes among the $k$ nodes with the highest scores; and the AUC, which evaluates the overall model's ability to distinguish asymptomatic from susceptible nodes. Both metrics are computed only on the evaluation pool (\( v \notin O(t_h) \)).

        The code used to train the GNN model and generate results is publicly available at \href{https://github.com/LIRA-UFRJ/Infected-Detection}{github.com/LIRA-UFRJ/Infected-Detection}. 

		
        

    \subsection{Observation Probability}
    \label{sec:result:obs_variation}
\begin{figure}[h]
        	\subfloat[BA model]{ \label{fig:auc_n3k_obsVar:BA}
        		\includegraphics[trim={4mm 0 3mm 0}, clip,width=0.475\textwidth]{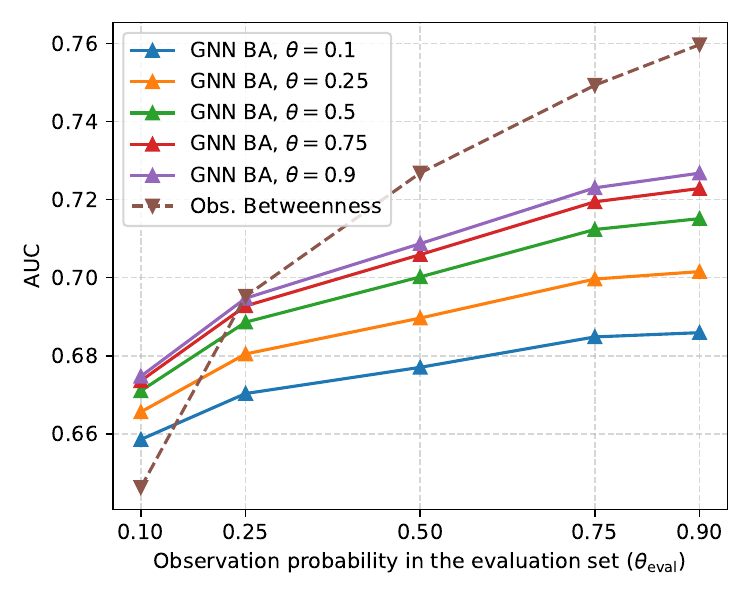}
        	}
        	\subfloat[WS model]{ \label{fig:auc_n3k_obsVar:WS}
        		\includegraphics[trim={4mm 0 3mm 0}, clip,width=0.475\textwidth]{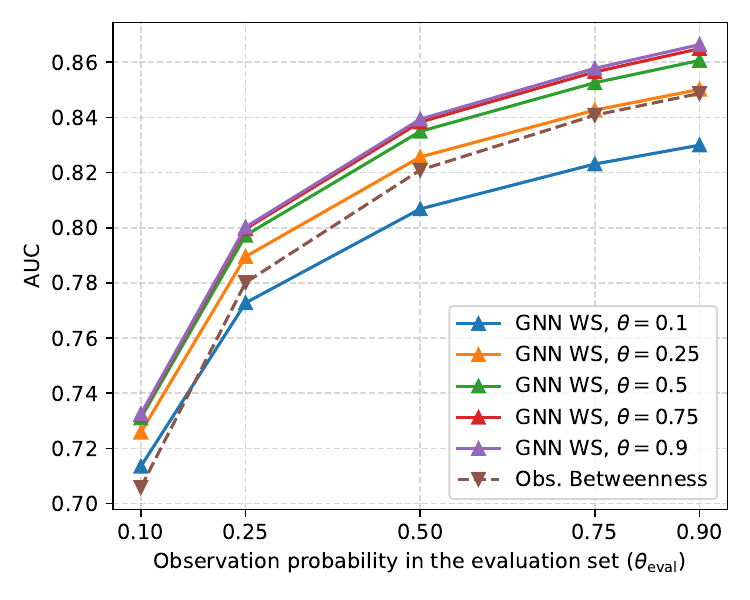}
        	}
        	\caption{
        		AUC performance (Y-axis) under different probabilities of observing an infected node (X-axis). Evaluation performed on networks with $3\mil$ nodes. The lines represent the mean AUC over the $1\mil$ samples in each dataset. 
        		Each full line corresponds to a trained GNN model performance (each one trained under a different observation probability), and the dashed line corresponds to the performance of the observed betweenness metric. 
        		Figure (a) shows the results for epidemics generated with the BA networks, and Figure (b) with the WS networks (training and testing conducted on the same network model). 
        	}
        	\label{fig:auc_n3k_obsVar}
        \end{figure}

        For the first set of experiments, the models (that were trained on networks of $3\mil$ nodes) are evaluated on test datasets generated under the same conditions ($3\mil$ nodes and the same network model). Results are shown in Figure \ref{fig:auc_n3k_obsVar}, which depicts the performance of the trained models for different observation probabilities $\theta$. 
        Note that each full line in the graphs is the performance of a GNN model, each one trained for a particular observation probability $\theta$, and tested under all $\theta_{\text{eval}}$ values considered.
        The dashed lines correspond to the performance of the observed betweenness metric \cite{Catarcione_2024}, which is also used as one of the input features of the GNN (as described in Section \ref{sec:metodo}). 

        It is notable that the results for the BA and WS network models are very different from each other. Figure \ref{fig:auc_n3k_obsVar:BA} shows that, for the BA model, the GNNs are only capable of surpassing the performance of the observed betweenness in epidemics that have a probability of observation $\theta_\text{eval} = 0.1$. 
        In contrast, Figure \ref{fig:auc_n3k_obsVar:WS} shows that, for WS networks, the GNNs present improved performance over the observed betweenness for all observation probabilities. Even when tested on a different $\theta_\text{eval}$ than the $\theta$ used for training, the GNNs trained with $\theta \geq 0.25$ show better performance on any of the evaluated $\theta_\text{eval}$ values. 

        It is understandable that WS networks are easier for the discrimination models since these networks have a much more regular structure than the BA networks. In a BA network, the epidemics spreading can behave very differently if the first infected node is a hub or if it is a peripheral node, for example, while in a WS network the epidemic spread should be similarly independent of the epidemic source. 
        Furthermore, due to the nature of BA networks, which tend to have one node that is attached to almost all other nodes in the network, this setting might be very difficult for GNNs, which rely on message exchange between nodes in order to compute their latent representation. With high degree hubs, it is likely that the nodes' representations are all becoming too homogeneous, even if the GNN has only two layers, which compromises its discrimination capacity. 
        Further study into this particular setting needs to be conducted in order to verify which models are most efficient for BA networks. 
        
        Interestingly, results show that it is worthwhile to train the model using a dataset with a greater observation probability $\theta$: on both network models, there is a monotonic increase in performance when training with higher observation rates, regardless of the $\theta_\text{eval}$ seen in the test dataset. 
        

	\subsection{Network Size}
    \label{sec:result:n_variation}
\begin{figure}[!h]
        	\centering
        	\includegraphics[trim={10mm 50mm 10mm 0}, clip,width=0.8\textwidth]{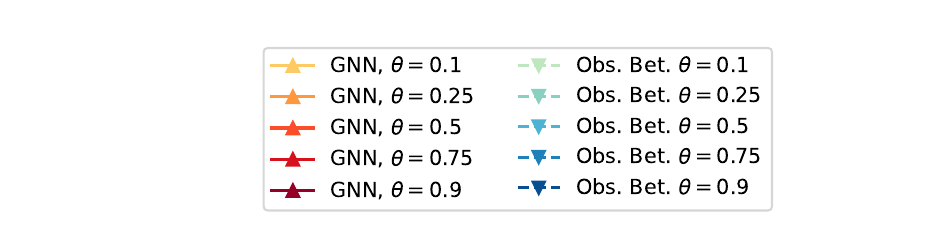}
        	\vskip -4mm
        	\subfloat[BA model]{ \label{fig:auc_nVar_obsTrain:BA}
        		\includegraphics[trim={3mm 0 3mm 0}, clip,width=0.48\textwidth]{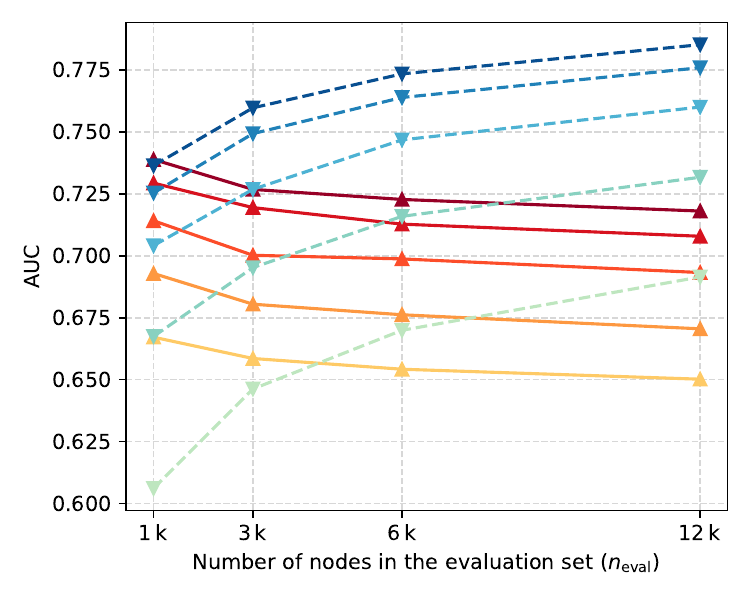}
        	} 
        	\subfloat[WS model]{ \label{fig:auc_nVar_obsTrain:WS}
        		\includegraphics[trim={3mm 0 3mm 0}, clip,width=0.48\textwidth]{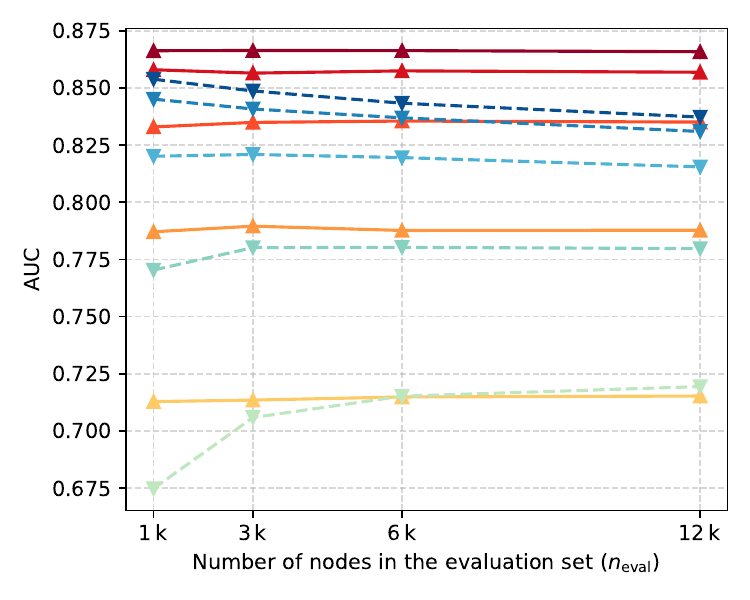}
        	}
        	\caption{
        		AUC performance (Y-axis) under different network sizes (number of nodes in the network -- X-axis). The evaluation is performed under the same observation probability of the GNN training ($\theta_\text{eval} = \theta$). 
        		The lines represent the mean AUC over the $1\mil$ samples in each dataset. 
        		Each full line corresponds to a trained GNN model performance (each one is trained under a different observation probability), and the dashed line corresponds to the performance of the observed betweenness metric. 
        		Figure (a) shows the results for epidemics generated with the BA networks, and Figure (b) with the WS networks (training and testing conducted on the same network model). 
        	}
        	\label{fig:auc_nVar_obsTrain}
        \end{figure}

    	For the second set of experiments, the effect of the network size is evaluated. In that sense, the GNNs, which were trained with instances of $3\mil$ nodes, are evaluated under datasets of $n_\text{eval} \in \{1\mil, 3\mil, 6\mil, 12\mil\}$ nodes. The results are shown in Figure \ref{fig:auc_nVar_obsTrain}. Note that the observation probability $\theta$ used during training is kept the same in the test datasets (all models are evaluated using $\theta_\text{eval} = \theta$).
        
        It is immediately apparent from Figure \ref{fig:auc_nVar_obsTrain} that the GNNs can extrapolate from the training data and are able to make informed predictions of the asymptomatic nodes in networks that are smaller and much larger than the ones seen during the GNN training. Especially for the WS model, the AUC performance of the GNNs suffers almost no change even when the instances have $12\mil$ nodes, thus four times larger than the ones used for training the GNN (which have $3\mil$ nodes). 
        
        Although results on BA networks show overall worse performance than the observed betweenness metric (as seen in the previous section), it is relatively stable with the increase of the test network size. It is also interesting to note that for smaller networks (that is, for networks with $1\mil$ nodes), the GNNs achieve better AUC performance than observed betweenness, and also when the observation probability is equal to 10\% ($\theta_{\text{eval}} = 0.1$). These are the instances where the observed betweenness has the worst performance, and it is remarkable that the GNNs are able to improve performance in such scenarios. 
        
\begin{figure}[h]
        	\centering
        	\includegraphics[trim={3mm 0 3mm 0}, clip,width=0.7\textwidth]{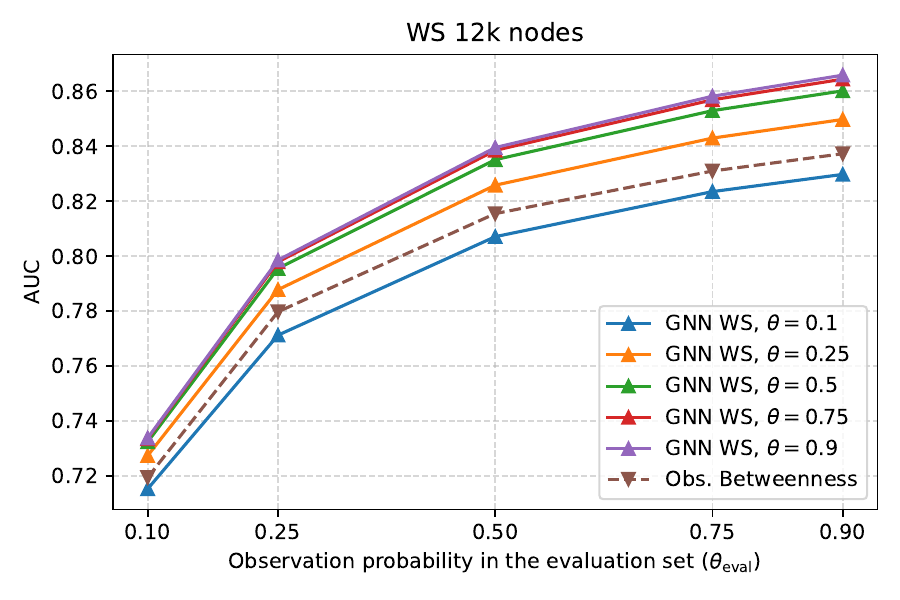}
        	\caption{
        		AUC performance (Y-axis) under different probabilities of observing an infected node (X-axis). Evaluation performed on $12\mil$ nodes WS networks. The lines represent the mean AUC over the $1\mil$ samples in each dataset. 
        		Each full line corresponds to a trained GNN model performance (each one is trained under a different observation probability), and the dashed line corresponds to the performance of the observed betweenness metric. 
        	}
        	\label{fig:auc_n12k_obsVar:WS}
        \end{figure}

        Another interesting result is that, when varying the observation probability of the test set ($\theta_{\text{eval}}$), the GNNs trained with $\theta = 0.9$ have the greatest overall performance. This phenomenon can be seen in Figure \ref{fig:auc_n3k_obsVar}, for test networks of $3\mil$ nodes, but also in Figure \ref{fig:auc_n12k_obsVar:WS}, which depicts the AUC performance of WS networks with $12\mil$ nodes. 
        
 		As seen before, increasing the observation probability of the training set $\theta$ increases the performance of the GNN prediction not only for its own observation probability, but also for test datasets with a smaller $\theta_{\text{eval}}$. 
 		Intuitively, inferring asymptomatic nodes for a higher observation gives more information on the epidemic spread, and thus the model is able to learn its patterns more efficiently. 
        
 		
 		Since it was observed that, for our experimental set, training the GNNs with a greater observation probability leads to a better overall performance, the subsequently presented results pertain only to GNNs trained with $\theta = 0.9$. 
        
    \subsection{Network Model}
    \label{sec:result:networkmodel_variation}
\begin{figure}[h]
    		\centering
    		\hspace*{5mm} \includegraphics[trim={0 13.3cm 2mm 0}, clip, width=0.8\textwidth]{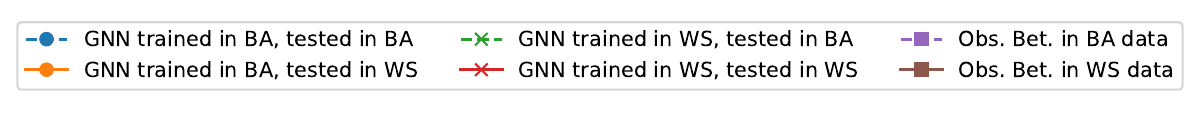}
    		\vskip -5mm
    		\subfloat[$\theta_{\text{eval}} = 0.1$]{ \label{fig:auc_nRede_nVar_obsTreino90:obs10}
    			\includegraphics[trim={3mm 0 3mm 11mm}, clip,width=0.475\textwidth]{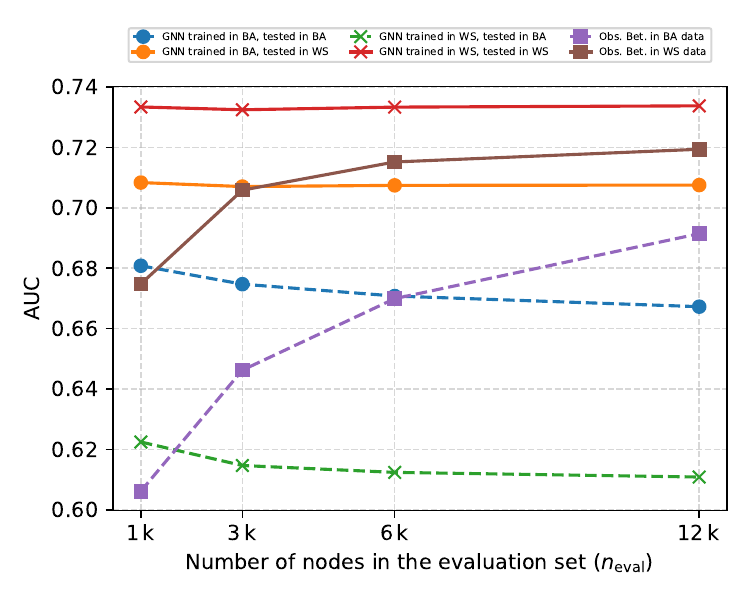}
    		}
    		\subfloat[$\theta_{\text{eval}} = 0.9$]{ \label{fig:auc_nRede_nVar_obsTreino90:obs90}
    			\includegraphics[trim={3mm 0 3mm 11mm}, clip,width=0.475\textwidth]{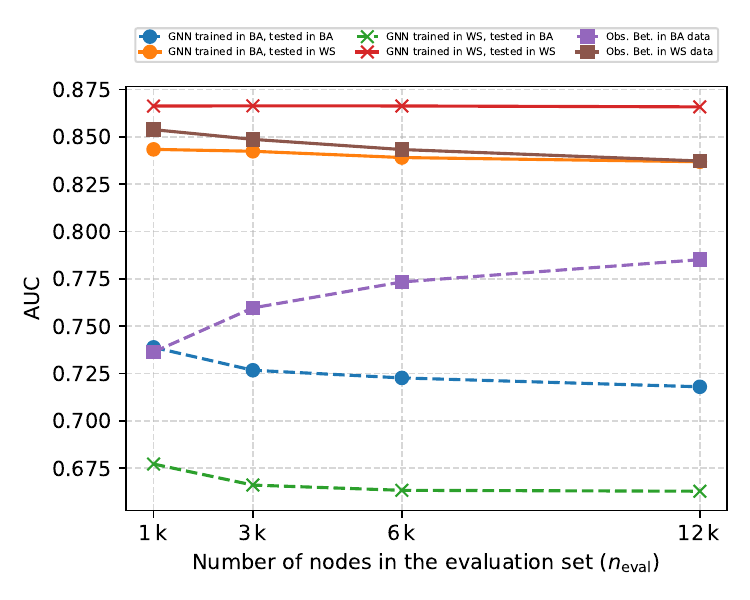}
    		}
    		\caption{
    			AUC performance (Y-axis) under different network sizes (number of nodes in the network -- X-axis). The lines represent the mean AUC over the $1\mil$ samples in each dataset. 
    			The GNNs used were trained with observation probability $\theta=0.9$. 
    			The full lines correspond to evaluations on BA data, and the dashed lines correspond to evaluations on the WS datasets. The circle markers stand for the GNN trained on BA data, the \enquote{X} markers stand for the GNN trained on WS, and the square markers to the observed betweenness metric. 
    			Figure (a) shows the results for epidemics with observation probability $\theta_\text{eval} = 0.1$, and Figure (b) with $\theta_\text{eval} = 0.9$. 
    		}
    		\label{fig:auc_nRede_nVar_obsTreino90}
    	\end{figure}

        The next set of experiments aims at evaluating the performance of the GNNs when test instances are from a different network model. In that sense, Figure \ref{fig:auc_nRede_nVar_obsTreino90} presents a comparison of the performance of the GNN models trained on BA and WS datasets and the performance of the observed betweenness metric by itself, when applied to test datasets from both network models.
    	
    	It is immediately apparent from Figure \ref{fig:auc_nRede_nVar_obsTreino90} that the performance of predictions conducted on the WS networks is generally better than the performance on BA networks, for all models. Also, one can note that the GNN model evaluated on the same network model it was trained on performs better than the model trained on a different network model. 
    	This is an expected result, which highlights the importance of the network model used in the training data. 
    	The training dataset is, in most cases, the key factor that enables the GNN to outperform the observed betweenness metric. 
        This is an important consideration for future extensions of the method to real-world networks: the structure of the network seen during training plays a critical role in the model's performance. 

    \subsection{Top-$k$ Analysis}
    \label{sec:result:topk}
        This set of experiments verifies whether the improved performance seen by the GNNs in the WS dataset under the AUC evaluation metric also holds for other performance metrics, namely the top-$k$ precision analysis. 
        This evaluation metric is often utilised in the literature and simply considers if the $k$ more likely nodes to be asymptomatic according to the model are indeed asymptomatic.

        Table \ref{tab:resultsWS_auc_topk} reports the top-$1\%$ and AUC performances of the GNN trained with WS data with $\theta=0.9$, and the observed betweenness metric by itself. 
        For these analyses, the number of nodes $k$ used is a percentage of the evaluation pool. In this instance, $k = 1\%$ of the evaluation pool is considered, which corresponds to an average of $ \left( 1 - 0.2 \,\theta_\text{eval} \right) \times 0.01 \, n_\text{eval} $ nodes evaluated. 
        Therefore, on $3\mil$-nodes networks with probability observation $\theta_\text{eval} = 0.1$, the top-$1\%$ metric evaluates $\approx 29.4$ nodes, and for $12\mil$ nodes with $\theta_\text{eval} = 0.9$, $k \approx 98.4$ nodes. 
    	
    	\begin{table}[!h] 
    		\centering
    		\caption{Comparison of AUC and top-$1\%$ performance of the GNN model (trained with WS networks with $\theta=0.9$) for test datasets with different numbers of nodes and observation probability $\theta_\text{eval}$. The results represent the mean performance and standard deviation.}
    		\label{tab:resultsWS_auc_topk}
            \vskip 2mm
    		\begin{tabular}{ccccccc}
    			\toprule
    			\phantom{\; 12k nodes \;} & \phantom{\; $\theta_{\text{eval}} = 0.1$ \;} & \phantom{\;\; top-1\% \;\;} & {\bfseries GNN} & \; &  {\bfseries Obs. Bet.} & \; \\
    			\toprule 
    			\multirow[c]{8}{*}{$12\mil$ nodes} & \multirow{3}{*}{$\theta_{\text{eval}} = 0.1$} & \color{tableLines} {\bfseries AUC} & \color{tableLines} $0.7338 \pm 0.0175$ & & \color{tableLines} $0.7194 \pm 0.0163$ & \; \\
                \cmidrule(l){3-6}
    			& & \color{tableLines2} {\bfseries top-1\%} & \color{tableLines2} $0.8438 \pm 0.0626$ & & \color{tableLines2} $0.7600 \pm 0.0592$ & \; \\
                %
                \cmidrule(l){2-7}
                & \multirow{3}{*}{$\theta_{\text{eval}} = 0.5$} & \color{tableLines} {\bfseries AUC} & \color{tableLines} $0.8395 \pm 0.0143$ & & \color{tableLines} $0.8155 \pm 0.0124$ & \; \\
                \cmidrule(l){3-6}
                & & \color{tableLines2} {\bfseries top-1\%} & \color{tableLines2} $0.9324 \pm 0.0404$ & & \color{tableLines2} $0.8280 \pm 0.0541$ & \; \\
    			\cmidrule(l){2-7}
    			& \multirow{3}{*}{$\theta_{\text{eval}} = 0.9$} & \color{tableLines} {\bfseries AUC} & \color{tableLines} $0.8659 \pm 0.0171$ & & \color{tableLines} $0.8373 \pm 0.0157$ & \; \\
                \cmidrule(l){3-6}
    			& & \color{tableLines2} {\bfseries top-1\%} & \color{tableLines2} $0.5246 \pm 0.0729$ & & \color{tableLines2} $0.3758 \pm 0.0600$ & \; \\
    			%
                \midrule
                \multirow[c]{8}{*}{$6\mil$ nodes} & \multirow{3}{*}{$\theta_{\text{eval}} = 0.1$} & \color{tableLines} {\bfseries AUC} & \color{tableLines} $0.7334 \pm 0.0206$ & & \color{tableLines} $0.7152 \pm 0.0193$ & \; \\
                \cmidrule(l){3-6}
    			& & \color{tableLines2} {\bfseries top-1\%} & \color{tableLines2} $0.8407 \pm 0.0784$ & & \color{tableLines2} $0.7672 \pm 0.0730$ & \; \\
                %
                \cmidrule(l){2-7}
                & \multirow{3}{*}{$\theta_{\text{eval}} = 0.5$} & \color{tableLines} {\bfseries AUC} & \color{tableLines} $0.8399 \pm 0.0159$ & & \color{tableLines} $0.8196 \pm 0.0138$ & \; \\
                \cmidrule(l){3-6}
                & & \color{tableLines2} {\bfseries top-1\%} & \color{tableLines2} $0.9298 \pm 0.0503$ & & \color{tableLines2} $0.8401 \pm 0.0647$ & \; \\
    			\cmidrule(l){2-7}
    			& \multirow{3}{*}{$\theta_{\text{eval}} = 0.9$} & \color{tableLines} {\bfseries AUC} & \color{tableLines} $0.8664 \pm 0.0204$ & & \color{tableLines} $0.8433 \pm 0.0193$ & \; \\
                \cmidrule(l){3-6}
    			& & \color{tableLines2} {\bfseries top-1\%} & \color{tableLines2} $0.5194 \pm 0.0890$ & & \color{tableLines2} $0.3899 \pm 0.0792$ & \; \\
                \midrule
                \multirow[c]{8}{*}{$3\mil$ nodes} & \multirow{3}{*}{$\theta_{\text{eval}} = 0.1$} & \color{tableLines} {\bfseries AUC} & \color{tableLines} $0.7325 \pm 0.0257$ & & \color{tableLines} $0.7059 \pm 0.0242$ & \; \\
                \cmidrule(l){3-6} 
    			& & \color{tableLines2} {\bfseries top-1\%} & \color{tableLines2} $0.8335 \pm 0.1055$ & & \color{tableLines2} $0.7670 \pm 0.0969$ & \; \\
                %
                \cmidrule(l){2-7}
                & \multirow{3}{*}{$\theta_{\text{eval}} = 0.5$} & \color{tableLines} {\bfseries AUC} & \color{tableLines} $0.8394 \pm 0.0194$ & & \color{tableLines} $0.8209 \pm 0.0162$ & \; \\
                \cmidrule(l){3-6}
                & & \color{tableLines2} {\bfseries top-1\%} & \color{tableLines2} $0.9309 \pm 0.0606$ & & \color{tableLines2} $0.8545 \pm 0.0808$ & \; \\
    			\cmidrule(l){2-7}
    			& \multirow{3}{*}{$\theta_{\text{eval}} = 0.9$} & \color{tableLines} {\bfseries AUC} & \color{tableLines} $0.8664 \pm 0.0267$ & & \color{tableLines} $0.8487 \pm 0.0241$ & \; \\ 
                \cmidrule(l){3-6}
    			& & \color{tableLines2} {\bfseries top-1\%} & \color{tableLines2} $0.5242 \pm 0.1168$ & & \color{tableLines2} $0.4110 \pm 0.1075$ & \; \\
                \midrule
                \multirow[c]{8}{*}{$1\mil$ nodes} & \multirow{3}{*}{$\theta_{\text{eval}} = 0.1$} & \color{tableLines} {\bfseries AUC} & \color{tableLines} $0.7334 \pm 0.0431$ & & \color{tableLines} $0.6748 \pm 0.0417$ & \; \\
                \cmidrule(l){3-6}
    			& & \color{tableLines2} {\bfseries top-1\%} & \color{tableLines2} $0.8176 \pm 0.1772$ & & \color{tableLines2} $0.7534 \pm 0.1623$ & \; \\
                \cmidrule(l){2-7}
                & \multirow{3}{*}{$\theta_{\text{eval}} = 0.5$} & \color{tableLines} {\bfseries AUC} & \color{tableLines} $0.8375 \pm 0.0270$ & & \color{tableLines} $0.8202 \pm 0.0243$ & \; \\
                \cmidrule(l){3-6}
                & & \color{tableLines2} {\bfseries top-1\%} & \color{tableLines2} $0.9188 \pm 0.1015$ & & \color{tableLines2} $0.8599 \pm 0.1209$ & \; \\
    			\cmidrule(l){2-7}
    			& \multirow{3}{*}{$\theta_{\text{eval}} = 0.9$} & \color{tableLines} {\bfseries AUC} & \color{tableLines} $0.8663 \pm 0.0453$ & & \color{tableLines} $0.8538 \pm 0.0395$ & \; \\
                \cmidrule(l){3-6}
    			& & \color{tableLines2} {\bfseries top-1\%} & \color{tableLines2} $0.5081 \pm 0.1909$ & & \color{tableLines2} $0.4289 \pm 0.1744$ & \; \\
                \bottomrule
    		\end{tabular}
    	\end{table}

        While the top-$1\%$ results show similar tendencies as AUC, in the sense that the GNN model is capable of enhancing the performance of the observed betweenness, it is interesting to note that the discrepancy is much higher than in the AUC results: for the top-$1\%$ evaluation metric, the performance increase with regards to the observed betweenness is always shown in the first decimal place. 
        Note that, for networks of $12\mil$ nodes and $\theta_\text{eval}=0.9$, the observed betweenness obtains an average of $0.38$ precision for the top-$1\%$ metric, but the GNN achieves an average of $0.52$. 

        While the results for $\theta_\text{eval} = 0.9$ might at first glance seem bad, $1\%$ of the evaluation pool corresponds to roughly half of all the asymptomatic nodes in the network. Therefore, this is a difficult scenario for any prediction model. 
        Note that the best top-$1\%$ performance is seen when $\theta_\text{eval} = 0.5$. This is a more favourable scenario because it has a better trade-off between the information the model has available (i.e., nodes observed as infected) and the number of asymptomatic nodes to classify (i.e., around half of the infected nodes).

%% file: conclusao.tex
\section{Conclusion}
\label{sec:conc}

    This work considered the problem of identifying asymptomatic nodes in an SI network epidemic using a single epidemic snapshot where a fraction of the infected nodes are not observed (the asymptomatic). 
    The performance of a Graph Neural Network (GNN) model with a set of node features computed from the snapshot is evaluated in this problem setting. 
    The proposed choice of features augments the classic epidemic state of a node and provides higher-level information for the GNN model.

    
    The GNN model was evaluated on epidemic datasets generated on Barabási-Albert (BA) and Watts-Strogatz (WS) network models with different network sizes and fractions of observed infected nodes. Results show that the GNN can generalise well with respect to the network size used during training. For the WS model, the GNN greatly outperformed the previously proposed observed betweenness metric.
    Results on BA networks are not as solid, indicating there is still room for future research and improvements in such scenarios. 
    Lastly, there is room for the conduction of several ablation studies, which could provide insights on how to improve further the promising results discussed in this work. 
    
     


